# Power and sample size for cluster randomized and stepped wedge trials: Comparing estimates obtained by applying design effects or by direct estimation in GLMM

## David M. Thompson

## Purpose

When observations are independent, formulae and software are readily available to plan and design studies of appropriate size and power to detect important associations. When observations are correlated or clustered, results obtained from the standard software require adjustment. This tutorial compares two approaches, using examples that illustrate various designs for both independent and clustered data.

One approach obtains initial estimates using software that assume independence among observations, then adjusts these estimates using a design effect (DE), also called a variance inflation factor (VIF). A second approach generates estimates using generalized linear mixed models (GLMM) that account directly for patterns of clustering and correlation.

The two approaches generally produce similar estimates and so validate one another. For certain clustered designs, small differences in power estimates emphasize the importance of specifying an alternative hypothesis in terms of means but also in terms of expected variances and covariances. Both approaches to power estimation are sensitive to assumptions concerning the structure or pattern of independence or correlation among clustered outcomes.

## Cluster randomized and stepped wedge designs

In a randomized controlled trial (RCT), investigators randomly assign individuals to one or more study arms, in which they experience different treatment. A cluster randomized trial (CRT) randomly assigns treatment not to individuals but to "clusters," that is, to clinics, physician practices or classrooms. Numerous individuals make up each cluster. Cluster randomization is useful when contamination is possible, for example, when a physician or group of physicians might find it difficult to adhere strictly to individual treatment assignments. When circumstances can make it hard for providers to administer different interventions to individuals under their care. treatment assignment by clinic permits all providers within a clinic to follow a single set of treatment guidelines and to offer care consistently to all individuals in a CRT. These advantages make cluster randomization a popular approach for health services research.

The stepped wedge design (SWD) also randomizes treatments by cluster. The SWD (Hussey and Hughes, 2007) can be an attractive alternative to the CRT for interventions whose efficacy is already supported. This is because the SWD is a type of crossover design where the crossover is "unidirectional." Individuals in each cluster experience the intervention following a non-intervention period of varying length. The design takes its name from the wedge-like appearance of depictions like Figure 1, which shows how the intervention is initiated in a series of steps.

**Figure 1. from Woertmann et al. (2013, Fig. 1, p.753)**

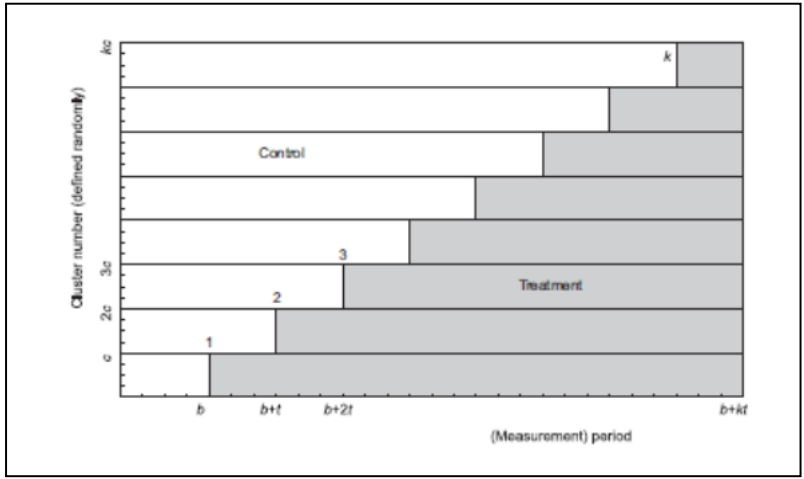

Both CRTs and SWDs randomly assign interventions not to individuals but according to a natural clustering unit such as the physician or the practice in which individuals receive care. As a result, both CRTs and SWDs require care in the calculation of power and sample size. Individuals who attend the same clinic are likely to be share characteristics that affect their propensity to respond to treatment. These shared within-cluster characteristics may relate to individuals inhabiting the same geographic area or community, or to commonalities in physician or practice styles (Thompson, Fernald, Mold, 2012, p. 235). Because they share important characteristics, individuals' responses may be more like those of others who frequent the same clinic than like those of individuals who receive care at other clinics. Calculations of power and sample size must account for this "clustering" or within-clinic correlation. The complexity of sample size calculations is even greater when individuals are followed over time and observed repeatedly.

## Two approaches for planning sample size and ensuring appropriate statistical power

This tutorial compares two approaches to planning sample size and ensuring appropriate levels of statistical power. The first approach employs a design effect (DE), also called a variance inflation factor (VIF). To apply this approach, the analyst first calculates a sample size estimate for a design that assumes independence among observations. Conventional software is widely available to estimate the sample size that affords acceptable power to detect, in a study with independent observations, a hypothesized effect.

To adjust this estimate for clustering or correlation among observations, investigators multiply the initial estimate by an appropriate design effect. The product predicts the size of the CRT or SWD that will afford the same statistical power as an RCT performed on unclustered observations (Donner & Klar, 2000, pp.112-113).

Useful explanations of formulae for DE are available for CRT (Teerenstra et al., 2012) and SWD (Woertman et al., 2013; de Hoop et al., 2015). Hooper et al. (2016) derive a general formula to calculate design effects for a broad range of designs with clustered and repeated observations. They show how formulae commonly used for the CRT and SWD are special cases of their general approach. These formulae require estimates of correlations among responses between individuals in the same cluster and, when individuals are measured more than once, among repeated measures on individuals.

The second approach, advocated by Littell et al. (2006), Stroup (1999, 2013) and Stroup et al. (2018, Chapter 14) directly estimates sample size and power using generalized linear mixed models (GLMM). The investigator initiates this approach by specifying an alternative hypothesis, defining the structure of the mean outcomes expected under that hypothesis, and then creating a dataset that exemplifies the expected means. In addition to creating an exemplary dataset with the expected means, the investigator must also specify expected values for variance components, and account for expected correlations or clustering among observed outcomes.

In a second step, the "exemplary dataset" is analyzed in a GLMM, which also accounts for the expected variance components, and which generates a test statistic. The examples of GLMM explored in this tutorial generate non-central F statistics. A third and final step calculates expected power using the test statistics. An Appendix to this tutorial provides background for this final step.

This tutorial demonstrates, for each of seven examples, how to arrive at estimates of power by applying a design effect, and also by estimating power directly in a GLMM. It shows how to construct exemplary datasets using SAS data steps, and how to construct GLMMs using SAS PROC MIXED or PROC GLIMMIX. To apply design effects in situations where observations are correlated, one must specify estimates for these correlations. Properly constructed GLMMs must include estimates of variance components that are equivalent to these correlations. The tutorial demonstrates how to arrive at these, whether the correlations relate to observations on individuals who are in the same cluster, or to repeated observations on the same individual.

## Comparing two approaches across seven examples of study designs

This tutorial examines seven study designs whose associated statistical models are of increasing complexity. In each example, alternative hypotheses are expressed in terms of group- and time-specific means for an outcome variable whose distribution is assumed to be Gaussian. The seven designs are:

    1. A two-arm RCT where the outcome is measured once, after a period of intervention.

    2. A two arm CRT where the outcome is measured once, after a period of intervention

3. A two arm RCT where the outcome is measured twice, before and after a period of intervention, in independent groups.

4. A two arm "cross-sectional" CRT where the outcome is measured twice, before and after a period of intervention, in independent groups.

5. A two arm "cohort" CRT where the outcome is measured twice, before and after a period of intervention, in groups comprised of the same individuals, whose responses are therefore correlated.

6. A "cross sectional" SWD where the outcome is measured repeatedly at each "step," some before and some after intervention, in clusters that recruit independent groups at each step.

7. A "cohort" SWD where the outcome is measuredly repeatedly at each "step," some before and some after intervention, in clusters that follow and observe the same individuals over time.

The tutorial uses a common outline to evaluate each design. First, it presents a statistical model that describes the structure of expected means and variances. It then demonstrates how to calculate and apply a design effect to estimate the sample size required to afford a desired level of statistical power. Next, it demonstrates in detail the construction of an exemplary dataset and companion GLMM to obtain a direct estimate of power. Finally, it compares power estimates obtained by either approach.

The two approaches generate similar results for several examples and, in doing so, validate one another. For other examples, small differences in results illustrate the approaches' sensitivity to the specification of the expected variance components.

# 1. A two-arm RCT where the outcome is measured once, after a period of intervention.

We consider first a study wherein the investigators hypothesize that the outcome of interest will be equal in two groups at the trial's outset but will differ by, on average, five points after a period of intervention. They also expect, perhaps on the basis of pilot data, that the outcome's true variance is 25 units (true standard deviation of 5 units) in both treatment arms. The investigators assume that randomization will ensure that the groups do not differ at baseline and so plan a single post-intervention measurement of the outcome.

*Statistical model, mean and variance structure*

The statistical model that formally describes this design is:

$$y_{ig} = \beta_0 + \beta_1 \text{arm} + e_i$$

The response for individual i in study arm g is a function of a single fixed effect. An indicator variable takes the value of g=0 for individuals in the arm receiving standard care and g=1 for those in the arm receiving the novel treatment. The model includes no random effects, only a within-subject residual or error $e_i$. Thus, the model's single variance component is:

$$\text{Var}(y_{ig}) = \text{Var}(\beta_0 + \beta_1 \text{arm} + e_i) = \sigma_e^2$$

Responses between individuals are independent. In a study with $N_t$ individuals, the matrix of expected variances and covariances (V) that describes this design is a square matrix with $N_t$ rows and columns. Elements on V's diagonal represent variances and are assumed all to equal $\sigma_e^2$. Off-diagonal elements are covariances which, in this design, are all equal to 0. Thus, the matrix of expected variances and covariances $V=\sigma_e^2 I_{N_t}$.

## *Applying conventional software to estimate sample size*

Conventional power and sample size software, such as SAS PROC POWER, shows that an RCT with two treatment arms requires group sizes of 17 (a total sample size of 34) to afford power of 0.807 to detect a hypothesized between-group difference of five units while assuming a standard deviation of 5 units in both groups. PROC POWER syntax is illustrated in Table 1.

## *Reproducing the results in a GLMM*

Littell et al. (2006), Stroup (1999, 2013) and Stroup et al. (2018, Chapter 14) explain the use of generalized linear mixed models (GLMM) to calculate power and sample size. Stroup (2013, pp. 469-470) presents the approach in three steps. The first step requires creation of an "exemplary" dataset that identifies groups, specifies group sizes, and states expected means under a specific alternative hypothesis. Table 2 illustrates a SAS data step that creates an exemplary dataset for this example. It specifies two groups, each of the size (n=17) specified earlier in the demonstration of PROC POWER, whose expected mean values for the outcome differ by 5 units.

In the second step, the exemplary dataset is analyzed, along with information on the expected variance components, in an appropriate statistical model. Table 1 illustrates, for this and for each example, the use of SAS PROC MIXED or PROC GLIMMIX to accomplish this step. Each example features a MODEL statement that reflects the statistical model. Each includes a PARMS statement that specifies the expected values for the variance components. The PARMS statement for this example specifies the expected value for the single variance component, $\sigma_e^2 = 25$. Other syntax prevents the procedure's usual practice of iterating to a solution.

A third and final step uses output from PROC MIXED or PROC GLIMMIX, including an F statistic and numerator and denominator degrees of freedom. An Appendix provides background for this step. Briefly, statistics output from the GLMM are used to calculate a non-central F statistic, which is compared to the critical value for the F distribution defined by a null hypothesis. The same SAS data step accomplishes this for all examples explored in this tutorial:

```
data power;
  set fstats;
  alpha=0.05;
  lambda=numdf*fvalue;                /*non-centrality parameter*/
  fcrit=finv(1-alpha, numdf, dendf, 0); /*critical value for F distrib under H0*/
  power= 1- probf (fcrit, numdf, dendf, lambda);
run;
```

*Comparing results from the two approaches*

Processing of output from GLMM in the DATA step above shows that the design affords power of 0.807 to detect the hypothesized between-arm difference of 5 units. This value is identical to the estimate generated by PROC POWER, and so the two approaches validate one another in this design (Table 1).

## 2. A two arm CRT where the outcome is measured once, after a period of intervention

The RCT described in the previous example randomly assigns individuals to treatment groups. The design would require an individual provider to treat individuals differently, according to their assigned treatment. Providers could find it difficult to accomplish this without "contamination" between treatment protocols. To permit each provider to deliver a single, consistent intervention, investigators can plan a CRT. They can recruit providers, and randomly assign each provider to follow just one of the treatment protocols under investigation. Then, each provider can execute this protocol for a group or cluster of individuals.

*Statistical model, mean and variance structure*

A statistical model that formally describes a CRT is:

$$y_{icg} = \beta_0 + \beta_1 arm + b_c + e_i \qquad (2)$$

The response for subject i, who receives care in cluster c, a cluster which is randomly assigned to treatment arm g, is a function of a single fixed effect (treatment arm). An indicator variable takes the value of g=0 for individuals in the arm receiving standard care and g=1 for those in the arm receiving the novel treatment. The model treats the cluster c as a random effect. Accordingly, the model implies two variance components:

$$Var(y_{icg}) = \sigma_y^2 = Var(\beta_0 + \beta_1 arm + b_c + e_i) = \sigma_c^2 + \sigma_e^2$$

The variance $\sigma_c^2$ is a between-cluster variance, while $\sigma_e^2$ is a between-subject or error variance. An intra-cluster correlation (ICC), symbolized as ρ, is defined from these two variances:

$$\rho = \frac{\sigma_c^2}{\sigma_c^2 + \sigma_e^2} = \frac{\sigma_c^2}{\sigma_y^2}$$

*Applying a design effect to estimate sample size*

To arrive at an appropriate sample size for the CRT, we can multiply, by an appropriate design effect (DE), the sample size estimate for the two-group RCT with no clustering. The resulting sample size will appropriately adjust a CRT so that it affords the same power (0.81) to detect a between-group difference of five points (assuming a total variance of 25).

For this two-arm design, the applicable design effect is:

$$DE = 1 + \rho(m - 1), \qquad (1)$$

where m represents the mean cluster size and ρ denotes the intraclass (or intracluster) correlation coefficient (ICC) defined above.

The DE is also called the variance inflation factor (VIF). In fact, formula (1) represents a factor that is the ratio of two variances. The DE illustrated in formula (1) is the ratio between an outcome's variance

(a) in a study with clusters whose average size is m: $m\sigma_c^2 + \sigma_e^2$ and

(b) in a study with no clustering, where each individual is treated as an independent cluster of size m=1: $\sigma_c^2 + \sigma_e^2$ (Bland, 2000; Donner and Klar, 2000; Thompson, Fernald & Mold, 2012).

This ratio is equal to $\frac{m\sigma_c^2 + \sigma_e^2}{\sigma_c^2 + \sigma_e^2} = \frac{m\sigma_c^2}{\sigma_c^2 + \sigma_e^2} + \frac{\sigma_e^2}{\sigma_c^2 + \sigma_e^2}$.

Substituting ρ for the quantity $\sigma_c^2/(\sigma_c^2 + \sigma_e^2)$, the ratio reduces to equation (1), the formula for the DE.,

$$\frac{m\sigma_c^2}{\sigma_c^2 + \sigma_e^2} + \frac{\sigma_e^2}{\sigma_c^2 + \sigma_e^2} = m\rho + (1-\rho) = 1 + \rho(m - 1).$$

To arrive at a valid design effect, investigators must specify a mean cluster size (m) and a realistic but appropriately conservative value for the ICC (ρ). We use, for all the examples that require it, an estimate for the ICC of 0.10 (Thompson, Fernald, Mold, 2012).

To plan a CRT with the same power as the RCT described, which has power of about 0.8 above to detect a between-group difference in the mean outcome of 5 units (while also assuming a common within-group variance of 25 units), we must inflate the RCT's estimated sample size of 17 individuals per treatment arm (or 34 total individuals). If the investigators regard as feasible a cluster size of six individals per provider cluster, the DE is 1+(6-1)*0.1 or 1.5. Multiplying the sample size estimated for the RCT (34 total individuals) by this design effect implies a total sample size for the CRT of 34*1.5 or 51. A CRT that recruits 10 providers, randomizes five of them to either treatment arm, then has each provider measure the post-intervention outcome in independent groups of 6 individuals, will involve 60 individuals and so should easily match the desired level of power. The researchers could recruit one less provider, randomizing four providers to one of the treatment arms and five to the other. This involves 9*6 or 54 individuals, still more than the predicted number needed to afford power of 0.8 under the hypothesized set of means and variance components.

## *Reproducing the results in a GLMM*

[Table 2](#) illustrates the SAS DATA step that produces an exemplary dataset for this example. The data set specifies the expected mean outcomes in nine provider clusters, each of which involve six individuals, where five providers are assigned to one treatment arm, and four are assigned to the other.

[Table 1](#) depicts the PROC GLIMMIX step used to analyze the exemplary data set. The procedure's MODEL statement reflects the statistical model defined above for the design. Its RANDOM statement defines cluster (or clinic) as a random effect, thus clarifying that observations on individuals in the same cluster are correlated. The PARMS statement provides values for the two variance components, dividing the total variance of 25 into a between-cluster (between-provider) variance $\sigma_c^2$ of 2.5 and a residual variance $\sigma_e^2$ of 22.5. The PARMS statement must list these values in the same sequence in which the SAS procedure reports "covariance parameter estimates." The user must review this carefully, because the sequence depends on whether one defines the variance components using a RANDOM or a REPEATED statement.

The analyst must also review the variance matrix (V) generated by PROC GLIMMIX (or PROC MIXED) to verify that the procedure has assigned the values for the variance components that the investigator expects and intends. In this example, the appropriate cluster-specific V matrix, for a clinic or cluster with six individuals, should exhibit a compound symmetric structure with a common variance of 25 and a between-cluster variance of $\rho\sigma_y^2$ or 2.5.

The GLIMMIX procedure employed in this example, which defines a random intercept for each clinic, induces the appropriate compound symmetric (or "exchangeable") structure in the clinic-specific V matrix:

| 25.0 | 2.5 | 2.5 | 2.5 | 2.5 | 2.5 |
| 2.5 | 25.0 | 2.5 | 2.5 | 2.5 | 2.5 |
| 2.5 | 2.5 | 25.0 | 2.5 | 2.5 | 2.5 |
| 2.5 | 2.5 | 2.5 | 25.0 | 2.5 | 2.5 |
| 2.5 | 2.5 | 2.5 | 2.5 | 25.0 | 2.5  |
| 2.5 | 2.5 | 2.5 | 2.5 | 2.5  | 25.0 |

The corresponding VCORR matrix verifies that the variance components chosen are consistent with a hypothesized ICC of 0.10.

|     |     |     |     |     |     |
|-----|-----|-----|-----|-----|-----|
| 1.0 | 0.1 | 0.1 | 0.1 | 0.1 | 0.1 |
| 0.1 | 1.0 | 0.1 | 0.1 | 0.1 | 0.1 |
| 0.1 | 0.1 | 1.0 | 0.1 | 0.1 | 0.1 |
| 0.1 | 0.1 | 0.1 | 1.0 | 0.1 | 0.1 |
| 0.1 | 0.1 | 0.1 | 0.1 | 1.0 | 0.1 |
| 0.1 | 0.1 | 0.1 | 0.1 | 0.1 | 1.0 |

The syntax that creates the GLMM for this example and others (Examples 4-7) defines one or more random intercepts. Models that employ random intercepts induce, as illustrated in the V and VCORR matrices above, covariance structures that are compound symmetric or "exchangeable." Hooper et al. (2016, p.4726) point out that "exchangeability assumptions are common to most existing approaches to sample size calculation for longitudinal cluster randomised trials." In this example, the "exchangeability assumption" is that the outcome's correlation is the same between any two individuals in the same cluster.

## *Comparing results from the two approaches*

The design effect of 1.5, applied to the sample size estimated for an RCT with unclustered data, suggests that a CRT with 51 individuals will afford power of 0.807 to detect the hypothesized between-arm difference in the outcome of five units. PROC GLIMMIX predicts that a CRT that involves nine clusters (5 randomly assigned to one treatment arm and 4 assigned to the other), each with six individuals, has a total sample size of 54 and affords power of 0.831. A CRT that involves 48 individuals (six in each of eight clusters) affords power of 0.788.

In this example, application of the design effect estimates a sample size that we cannot match in a GLMM using clusters of equal size. Consequently, we cannot produce the matching results that would permit the methods to validate one another. The GLMM offers the flexibility to consider scenarios with different numbers of clusters, and different cluster sizes. By permitting some clusters to have six individuals and some seven, we can construct an exemplary dataset with eight clusters and exactly 51 individuals, the sample size required according to application of the DE. PROC GLIMMIX predicts that a CRT that involves 51 individuals (with six or seven individuals in each of eight clusters) affords power of 0.803.

This estimate is close to the power of 0.807 predicted using PROC POWER. Even so, it does not constitute a validation because the design's average cluster size differs from the value of six assumed in the calculation of the design effect. The example illustrates, however, an advantage for the GLMM's flexibility in accommodating variety in the number and size of clusters. Practical considerations related to clinic and individual recruitment could require this kind of flexibility.

## 3. A two arm RCT where the outcome is measured twice, before and after a period of intervention, in independent groups.

The tutorial's first two examples involve designs that measure an outcome once, following an intervention, on individuals in two treatment arms. This example addresses a "pre-post" design. Pre-post designs have certain advantages. By measuring the outcome in both arms of a study prior to intervention, investigators can assess the success of randomization and ensure that the groups are comparable at baseline. The pre-post design also permits comparison of the change in the outcome over time. By focusing on a "difference in differences," it accounts for the possibility that both groups may demonstrate change, but still determine whether the degree of change differs between groups.

*Statistical model, mean and variance structure*

The statistical model that describes sources that affect the response for individual i, at time t, within treatment arm g, is:

$$y_{itg} = \beta_0 + \beta_1 arm + \beta_2 time + \beta_3 (arm * time) + e_i$$

As in previous examples, an indicator variable *arm* takes the values of g=0 for the group receiving standard care and g=1 for the group receiving the novel treatment. The indicator variable *time* takes the value t=0 for the baseline measurement and t=1 for the post-intervention assessment.

Parameters associated with fixed effects are denoted by β. $\beta_0$ estimates baseline responses for individuals randomly assigned to receive standard care, while $\beta_1$ estimates the difference, at baseline, between mean responses among individuals in the study's two arms. $\beta_2$ estimates the mean change in the outcome, between the two measurements, *among individuals who receive standard care*.

$\beta_3$ estimates a "difference in differences;" it compares change over time among individuals receiving the novel treatment with those receiving standard care. The coefficient $\beta_3$ quantifies the treatment*time interaction, which is a critical comparison in this pre-post design,. The estimate of $\beta_3$ permits assessment of whether change in the group receiving the novel treatment differs from change in the group receiving standard care. The coefficient's estimate can detect, for example, whether those who receive the novel treatment improve more than those receiving standard care.

Groups of individuals who are observed before and after intervention are distinct and independent in either treatment arm. Therefore, like example 1, the statistical model contains no random effects. It implies a single variance component, a within-subject variance $\sigma_e^2$. Because responses on all individuals are presumed to be independent, the V matrix is of the form $\sigma_e^2 I_{N_t}$.

Figure 2 illustrates a hypothetical scenario in which both groups' mean scores are 54 on the baseline measurement, where one group improves by two units after intervention, and the other group improves by seven units. As in the previous examples, the magnitude of the hypothesized between-group effect, defined now as a "difference in differences," is five units. As in previous examples, the outcome's expected total variance is 25 units.

**Figure 2. Hypothesized mean structure for a design with two arms and two times of measurement.**

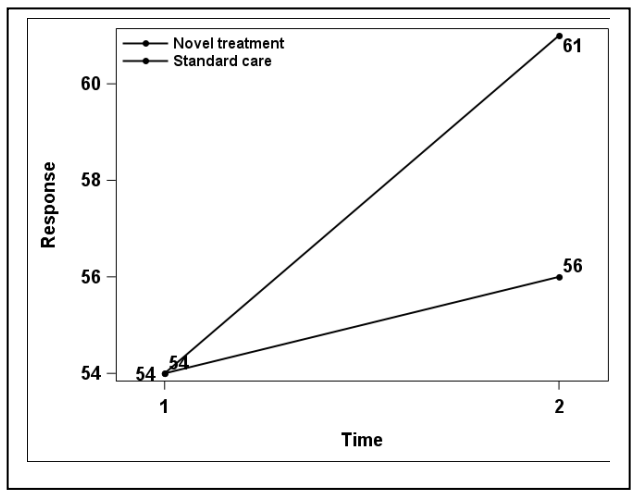

Conventional sample size software, such as SAS PROC GLMPOWER, shows that an RCT with two treatment arms, must enroll a total of 128 individuals to afford power of 0.789 to detect superior improvement among those in the group receiving novel treatment (Table 1). This implies that, in either treatment arm, 32 individuals are measured at baseline, and independent groups of 32 individuals are measured after intervention.

## *Reproducing the results in a GLMM*

Table 2 shows the steps that create an exemplary dataset for this pre-post RCT with independent groups.
Table 1 illustrates the PROC GLIMMIX step that creates a GLMM appropriate for this study design.

*Comparing results from the two approaches*

SAS PROC GLMPOWER estimates the design's power to detect the expected treatment*time interaction at 0.789. The GLMM constructed using SAS PROC GLIMMIX calculates the design's power to detect the hypothesized treatment*time interaction to be 0.801. The similarity of results is again evidence that the two approaches validate one another.

## 4. A two arm "cross-sectional" CRT where the outcome is measured twice, before and after a period of intervention, in independent groups.

The RCT described in Example 3 could require individual providers to treat individuals differently on the basis of their random assignment to a treatment. To permit providers to deliver a single, consistent intervention to all individuals, and to eliminate the threat of contamination between protocols, investigators can instead plan a CRT. They can recruit providers, randomly assign each of them to a treatment arm, then ask each to follow the same treatment protocol for a group or cluster of individuals.

A CRT that measures the outcome at baseline (prior to intervention), and again following the intervention, can be cross-sectional; it can measure the outcome on independent groups of individuals at either time point. Alternatively, each cluster can be composed of an intact cohort of individuals, whom the provider follows and in whom the outcome is measured at both time points. Example 4 describes a cross-sectional CRT, while Example 5 describes a cohort CRT.

*Statistical model, mean and variance structure*

A statistical model that captures the structure of expected means and variances in the cross-sectional CRT design is:

$$y_{ictg} = \beta_0 + \beta_1 arm + \beta_2 time + \beta_3 (arm * time) + b_c + b_{ct} + e_i$$

The model includes the fixed and the random effects that contribute to the value of the outcome y for subject i in clinic or cluster c at time t in arm g. The indicator variable *arm* takes the values of g=0 for the group receiving standard care and g=1 for the group receiving the novel treatment. The indicator variable *time* takes the value t=0 for the baseline measurement and t=1 for the post-intervention assessment.

The model's subscripts, and the notation for its random effects, are motivated by a model published by Teerenstra et al. (2012, Equation 1). They employ the model to provide a sample size formula for planning a CRT "that is analyzed using ANCOVA with the outcome at baseline as a covariate" (Teerenstra et al., 2012, p. 2170).

Consistent with other examples, $\beta_0$ estimates baseline responses for individuals in clusters randomly assigned to receive standard care, while $\beta_1$ estimates the difference, at baseline, between responses in clusters in the study's two arms. $\beta_2$ estimates the change, between the two measurements, in clusters receiving standard care. $\beta_3$ estimates a "difference in differences;" it compares change over time in clusters receiving the novel treatment with those receiving standard care.

Among the model's random effects, the intercept $b_{ct}$ quantifies variations in the outcome around the arm-specific mean in cluster c *at a given time of measurement*. The random intercept $b_c$ applies to observations obtained within a cluster *regardless of the time of measurement*.

This model divides the total variance among observed outcomes into three components:

$$\text{Var}(y_{ictg}) = \sigma_y^2 = V[\beta_0 + \beta_1 \text{arm} + \beta_2 \text{time} + \beta_3(\text{arm} * \text{time}) + b_c + b_{ct} + e_i] = \sigma_c^2 + \sigma_{ct}^2 + \sigma_e^2$$

The model's inclusion of the random effects $b_c$ and $b_{ct}$ and their corresponding variances $\sigma_c^2$ and $\sigma_{ct}^2$ is important because individuals in the same clinic are likely to exhibit commonalities in their responses whether they are measured at the same time or at different times. One might assume that, in a design like this, in which different groups of individuals are observed at different times, the responses might be uncorrelated. However, individuals who attend the same clinic share characteristics that can produce similar propensities to respond to treatment, even when they are measured at different times. Teerenstra et al. (2012) refer to this property as cluster autocorrelation ($\rho_c$).

Cluster autocorrelation ($\rho_c$), along with the ICC ($\rho$), quantify correlations between measurements obtained on different members of a cluster at the same and at different time points. For this design, Teerenstra et al. (2012) define:

$$\rho_c = \frac{\sigma_c^2}{\sigma_c^2 + \sigma_{ct}^2} \quad , \quad \rho = \frac{\sigma_c^2 + \sigma_{ct}^2}{\sigma_c^2 + \sigma_{ct}^2 + \sigma_e^2} \quad , \text{ so that } \rho * \rho_c = \frac{\sigma_c^2}{\sigma_c^2 + \sigma_{ct}^2 + \sigma_e^2}$$

Analysts may assume that cluster autocorrelation, $\rho_c$ is equal to 1 and, equivalently, that $\sigma_{ct}^2 = 0$. This implies that measurements on individuals within a cluster are equally correlated whether they are measured at the same time or at different times. Analysts can choose a value for $\rho_c$ that is less than one when they expect or hypothesize that observations obtained on individuals in the same cluster at different times are less correlated than observations obtained at the same time (Hooper et al, 2016, p.4727).

In this design, observations are obtained on the n members of a cluster at either of two time points. Therefore, the V matrix for a cluster with n individuals will have 2n rows and 2n columns. The V matrix is composed of four nxn submatrices:

$$\begin{matrix} A & B \\ B & A \end{matrix}$$

The submatrices A are a nxn and illustrate variances and covariances among the cluster's n observations at baseline and at the post-intervention measurement. Values on the main diagonal of A are all equal to the total variance, $\sigma_y^2$.

The submatrix A's off-diagonal elements, which are covariances between measurements obtained on individuals within the same cluster at the same time, are all equal to $\rho\sigma_y^2$. To understand this result, we review Teerenstra's definition for the ICC (Teerenstra et al., 2012, Equation 6). They remark that the ICC describes the "correlation at one time point in the same cluster between two different subjects" (Teerenstra et al., 2012, p. 2171). For this example, if we assume that within-subject variance is the same in all clusters, then where i≠i`, that is, for two different individuals in the same cluster, the ICC is:

$$\rho = \frac{\sigma_c^2 + \sigma_{ct}^2}{\sigma_c^2 + \sigma_{ct}^2 + \sigma_e^2} = \frac{\text{Cov}(y_{it}, y_{i`t})}{(\text{Var } y_{it})^{1/2}(\text{Var } y_{i`t})^{1/2}} = \frac{\text{Cov}(y_{it}, y_{i`t})}{\sigma_y^2}$$

Therefore, the covariances $\text{Cov}(y_{it}, y_{i`t})$ in the off-diagonal elements of submatrix A are equal to $\rho\sigma_y^2$.

The submatrices B represent covariances between observations obtained on different individuals at different times. All elements in B are equal to $\rho_c\rho\sigma_y^2$. This result follows from Teerenstra et al. (2012, Equation (5)), which defines r, "the correlation between a cluster mean at baseline … and at follow-up" (p. 2171) as:

$$r = \frac{n\rho}{1+(n-1)\rho}\rho_c + \frac{1-\rho}{1+(n-1)\rho}\rho_s$$

where n is the cluster size and $\rho_c$ is the coefficient for cluster autocorrelation previously discussed. The coefficient $\rho_s$ quantifies subject autocorrelation, which will be defined in Example 5. In this example, because different groups of individuals are measured at baseline and at follow-up in each cluster, subject autocorrelation ($\rho_s$) is equal to zero.

The formula for r relates to cluster means. To arrive at a correlation for individuals within a cluster, consider n=1. Thus, the correlation between individual observations within a cluster at baseline and follow up is equal to $\rho_c * \rho$. The covariances that make up the nxn elements of B are all equal to $\rho_c\rho\sigma_y^2$.

Having arrived at these expected values for elements of the V matrix, we can verify that the GLMM faithfully produces them when we analyze an exemplary dataset that is appropriate for this design. Before we demonstrate how to estimate sample size and power in a GLMM, we will demonstrate how to arrive at and apply an appropriate design for this study design. Then, we will apply the design effect to the sample size estimate obtained in Example 3 using PROC GLMPOWER.

### *Applying a design effect to estimate sample size*

Teerenstra et al. (2012, pp.2171-2172) advise: "First calculate the sample size for a cluster randomized trial that is analyzed on the follow-up scores (i.e. multiply the sample size [estimated] according to [an independent] t-test on the follow-up scores with the factor $[1 + (n -$ 1)ρ]. Then multiply this number with the design effect $(1 - r^2)$." Rutterford, Copas and Eldridge (2015, Equation 23, p.1059) report the equivalent design effect for this pre-post design as:

$$DE = [1 + (n - 1)\rho] * (1 - r^2) \qquad (2)$$

where r, as defined earlier, is equal to $\frac{n\rho}{1+(n-1)\rho}\rho_c + \frac{1-\rho}{1+(n-1)\rho}\rho_s$

Hooper et al. (2016, p.4722) use different notation but formulate an equivalent design effect for this "repeated cross-section cluster randomised trial with an ANCOVA design." Hooper et al. (2016, pp. 4725-4726) call Equation (2) an "overall design effect" and define it as "the product of the design effect due to repeated assessment $[1 - r^2]$ and the design effect due to cluster randomizing $[1 + (n - 1)\rho]$ ..."

Cluster autocorrelation, $\rho_c$, is discussed above. Subject autocorrelation, $\rho_s$, the "correlation between baseline and follow-up [measures on] subjects" within a cluster (Teerenstra et al., 2012, p.2170), will be more fully discussed in the next example. For this pre-post design, which involves two within-cluster measurements that are repeated on independent groups, the subject autocorrelation $\rho_s$ is equal to zero.

In applying the design effect to this example, we maintain the assumptions for the cluster size, n=10, and the ICC, ρ=0.10. We add the assumption that cluster autocorrelation, $\rho_c = 0.4$, so that the design effect is:

$$DE = [1 + (n - 1)\rho] * \left(1 - \left[\frac{n\rho}{1 + (n - 1)\rho}\rho_c\right]^2\right) = 1.9 * (1 - \left[\frac{(10)(0.1)(0.4)}{1.9}\right]^2) = 1.816$$

Multiplying the total sample size estimated for an RCT, 128, by the design effect of 1.816 yields a sample estimate for the CRT of 232.4.

This suggests that a CRT with a total of 240 individuals, with 20 individuals (10 measured before and 10 measured after the intervention) in each of 12 clinic clusters should afford power of at least 0.8 to detect the hypothesized group*time interaction.

### *Reproducing the results in a GLMM*

Table 2 contains a DATA step that creates an exemplary dataset that involves 240 individuals in two treatment arms. The dataset's arm- and time-specific mean responses reflect the hypothesized between-group differences depicted in Figure 2.

To construct a GLMM that accommodates this multilevel design (individuals nested within clusters), we must specify values for the variance components $\sigma_c^2$ and $\sigma_{ct}^2$.

$\rho = 0.1 = \frac{\sigma_c^2 + \sigma_{ct}^2}{\sigma_c^2 + \sigma_{ct}^2 + \sigma_e^2} = \frac{\sigma_c^2 + \sigma_{ct}^2}{25}$ . Therefore, $\sigma_c^2 + \sigma_{ct}^2 = 2.5$

$\rho_c = 0.4 = \frac{\sigma_c^2}{\sigma_c^2 + \sigma_{ct}^2} = \frac{\sigma_c^2}{2.5}$. Therefore, $\sigma_c^2 = 1$ and $\sigma_{ct}^2 = 1.5$

Because the total variance $\sigma_y^2 = 25$, it follows that $\sigma_e^2 = 22.5$

The PROC MIXED step that constructs the GLMM appropriate for this example (Table 1) incorporates two RANDOM statements that define random intercepts. The PARMS statement specifies values for the corresponding covariance parameters.

Construction of the GLMM for this example requires two modifications in the use of SAS procedures. First, it employs PROC MIXED instead of PROC GLIMMIX. Stroup et al. (2018) point out that, in models with relatively complex variance structures, "an artifact of the PROC GLIMMIX algorithm" can cause it to return a warning that "values given in PARMS statement are not feasible" (Stroup et al., 2018, Sections 4.4.4 and 14.5.2). They suggest strategies for altering the values that are included in the PARMS statement. Ultimately, because PROC MIXED and PROC GLIMMIX should produce identical results when applied to Gaussian responses, they suggest using PROC MIXED when one encounters the problem.

The second modification relates to the method for determining the denominator degrees of freedom (DDF), which are used in the calculation of the F statistics that are central to the determination of power. The previous examples have relied on the procedures' default methods establishing DDF. However, different methods produce values for the DDF "when the effect being tested is an interaction between two variables measured at different levels of the hierarchy" (Kupzyk, 2011, p. 31). That is precisely the case in this example, where the group*time interaction is of primary interest. The default methods for calculating DDF will be anti-conservative and will overestimate power. Stroup et al. (2018, Sect. 8.2.3) state that "the Kenward-Roger adjustment … should be considered mandatory standard operating procedure when analyzing repeated measures data." However, the Kenward-Roger, as well as the Satterthwaite adjustment, requires calculation of an asymptotic variance matrix. Because we have included syntax that prevents the PROC MIXED or PROC GLIMMIX steps from iterating to solutions, this matrix is not calculated. Consequently, the examples in this tutorial use, when appropriate, the Between-Within method to determine denominator degrees of freedom. The method chooses an appropriately conservative value for the DDF. Some researchers advocate general use of the Between-Within method for binary outcomes (Li and Redden, 2015).

As in previous examples, the investigator must inspect the V matrix generated by the GLMM to verify that it faithfully reflects the expected values for the variance components, The GLMM generates a cluster-specific V matrix composed of four submatrices, each with ten rows and ten columns. These are the appropriate dimensions for clusters whose ten individuals are each observed twice.

$$V = \begin{matrix} A & B \\ B & A \end{matrix}$$

The submatrices A demonstrate the intended form. Values on the main diagonal reflect the total variance, $\sigma_y^2 = 25$. The off-diagonal elements are covariances equal to $\rho\sigma_y^2 = (0.1)(25) = 2.5$.

| 25.0 | 2.5 | 2.5 | 2.5 | 2.5 | 2.5 | 2.5 | 2.5 | 2.5 | 2.5 |
|---|---|---|---|---|---|---|---|---|---|
| 2.5 | 25.0 | 2.5 | 2.5 | 2.5 | 2.5 | 2.5 | 2.5 | 2.5 | 2.5 |
| 2.5 | 2.5 | 25.0 | 2.5 | 2.5 | 2.5 | 2.5 | 2.5 | 2.5 | 2.5 |
| 2.5 | 2.5 | 2.5 | 25.0 | 2.5 | 2.5 | 2.5 | 2.5 | 2.5 | 2.5 |
| 2.5 | 2.5 | 2.5 | 2.5 | 25.0 | 2.5 | 2.5 | 2.5 | 2.5 | 2.5 |
| 2.5 | 2.5 | 2.5 | 2.5 | 2.5 | 25.0 | 2.5 | 2.5 | 2.5 | 2.5 |
| 2.5 | 2.5 | 2.5 | 2.5 | 2.5 | 2.5 | 25.0 | 2.5 | 2.5 | 2.5 |
| 2.5 | 2.5 | 2.5 | 2.5 | 2.5 | 2.5 | 2.5 | 25.0 | 2.5 | 2.5 |
| 2.5 | 2.5 | 2.5 | 2.5 | 2.5 | 2.5 | 2.5 | 2.5 | 25.0 | 2.5 |
| 2.5 | 2.5 | 2.5 | 2.5 | 2.5 | 2.5 | 2.5 | 2.5 | 2.5 | 25.0 |

The submatrices B also display the intended form, appropriately reflecting cluster autocorrelation of $\rho_c = 0.4$, such that their elements are all equal to $\rho_c \rho \sigma_y^2 = (0.4)(0.1)(25) = 1$.

| 1.0 | 1.0 | 1.0 | 1.0 | 1.0 | 1.0 | 1.0 | 1.0 | 1.0 | 1.0 |
|---|---|---|---|---|---|---|---|---|---|
| 1.0 | 1.0 | 1.0 | 1.0 | 1.0 | 1.0 | 1.0 | 1.0 | 1.0 | 1.0 |
| 1.0 | 1.0 | 1.0 | 1.0 | 1.0 | 1.0 | 1.0 | 1.0 | 1.0 | 1.0 |
| 1.0 | 1.0 | 1.0 | 1.0 | 1.0 | 1.0 | 1.0 | 1.0 | 1.0 | 1.0 |
| 1.0 | 1.0 | 1.0 | 1.0 | 1.0 | 1.0 | 1.0 | 1.0 | 1.0 | 1.0 |
| 1.0 | 1.0 | 1.0 | 1.0 | 1.0 | 1.0 | 1.0 | 1.0 | 1.0 | 1.0 |
| 1.0 | 1.0 | 1.0 | 1.0 | 1.0 | 1.0 | 1.0 | 1.0 | 1.0 | 1.0 |
| 1.0 | 1.0 | 1.0 | 1.0 | 1.0 | 1.0 | 1.0 | 1.0 | 1.0 | 1.0 |
| 1.0 | 1.0 | 1.0 | 1.0 | 1.0 | 1.0 | 1.0 | 1.0 | 1.0 | 1.0 |
| 1.0 | 1.0 | 1.0 | 1.0 | 1.0 | 1.0 | 1.0 | 1.0 | 1.0 | 1.0 |

### *Comparing results from the two approaches*

Application of the design effect (1.816) predicts that a sample of 232.4 individuals will afford power of 0.789 (the value estimated using PROC GLMPOWER) to detect the hypothesized between-arm differences in outcome. Table 1 shows that the GLMM, constructed to involve 240 (in 12 clusters with 20 individuals per cluster), affords power of 0.813. Application of a design effect proposed by Teerenstra et al. (2012), and the use of a carefully constructed GLMM, produce similar predictions of power.

The example incidentally illustrates an advantage for carefully considering the magnitude of cluster autocorrelation. Analysts with limited information may simply propose a value for $\rho_c$ of 1. Hooper et al. (2016, p. 4727) caution investigators against this practice. Doing so "overestimate[s] the correlation between sample means from the same cluster at different times" and "hence, under-estimate[s] the required sample size." Were we to assume cluster autocorrelation of 1 for this example, we would apply a design effect of 1.373, which is substantially lower than the value of 1.816 for the DE calculated under the assumption that $\rho_c = 0.4$. Applying the lower DE would underestimate the required sample size and produce an underpowered study.

## 5. A two arm "cohort" CRT where the outcome is measured twice, before and after a period of intervention, in the same individuals whose responses are therefore correlated.

The previous example, for which participating clinics recruit independent and unique groups for measurement at two times, is plausible but perhaps not preferred. A more powerful and more persuasive design would track the same group of individuals at each clinic and compare their responses before and after intervention.

### *Statistical model, mean and variance structure*

A statistical model appropriate for the cohort CRT design is:

$$y_{ictg} = \beta_0 + \beta_1 arm + \beta_2 time + \beta_3 (arm * time) + b_c + b_{ct} + b_s + b_{st} \quad (9)$$

Like models described for earlier examples, this one lists the fixed effects (denoted by β) and random effects (denoted by b) that contribute to the value of the outcome y for subject i in clinic or cluster c at time t in arm g. Notation for the four random effects is motivated, as before, by Teerenstra et al. (2012). Two of the four random intercepts, $b_c$ and $b_{ct}$, were described in connection with Example 4. The intercept $b_{ct}$ quantifies variation, around the arm-specific mean, of responses observed within clusters assigned to that treatment *at a given time of measurement*. The random intercept $b_c$ applies to observations obtained within a cluster *regardless of the time of measurement*.

This model additionally specifies the individual as a random effect. Providers in each cluster follow the same cohort of individuals over time and measure them twice. The random intercept $b_{st}$ quantifies variation, around a cluster-specific mean, for responses observed in individuals in that cluster *at a given time of measurement*. The random intercept $b_s$ applies to observation obtained on individuals within that cluster *regardless of the time of measurement*.

This model identifies four variance components:

$$Var(y_{ictg}) = var\{\beta_0 + \beta_1 arm + \beta_2 time + \beta_3(arm * time) + b_c + b_{ct} + b_s + b_{st}]$$

$$\sigma_y^2 = \sigma_c^2 + \sigma_{ct}^2 + \sigma_{st}^2 + \sigma_s^2$$

The previous example described how the variance components $\sigma_c^2$ and $\sigma_{ct}^2$ relate to cluster autocorrelation ($\rho_c$). The variance components $\sigma_{st}^2$ and $\sigma_s^2$ relate to subject autocorrelation ($\rho_s$), which [Teerenstra et al.](#) (2012, p. 2170) describe as the "correlation between baseline and follow-up [measures on] subjects within a cluster." Teerenstra et al (2012, Equation 2) define it:

$$\rho_s = \frac{\sigma_s^2}{\sigma_s^2 + \sigma_{st}^2}$$

in terms of $\sigma_{st}^2$, the outcome's variance within individuals observed at a given time of measurement, and of $\sigma_s^2$, the outcome's variance within individuals regardless of the time measurement.

A cluster-specific V matrix that appropriately displays variances and covariances among observations obtained in this design will have 2n rows and columns, where n is the number of individuals within a cluster. For example, the 6x6 V matrix for a cluster with n=3 individuals is comprised of an arrangement of 2x2 submatrices:

$$\begin{matrix} A & B & B \\ B & A & B \\ B & B & A \end{matrix}$$

The 2x2 submatrices A describe variances and covariances between paired measurements obtained on a single individual. The number of these submatrices, arranged on V's diagonal, reflects the number of individuals in the cluster. The 2x2 submatrices B describe covariances between observations on pairs of different individuals (i≠i`) within the same cluster.

As in previous examples the diagonal elements of each submatrix A are simply the expected variances of the outcome $\sigma_y^2$. The expected values for the off-diagonal elements in A reflect r, "the correlation between a cluster mean at baseline … and at follow-up" (Teerenstra et al., 2102, Equation 5):

$$r = \frac{n\rho}{1+(n-1)\rho}\rho_c + \frac{1-\rho}{1+(n-1)\rho}\rho_s$$

where ρ, the ICC, describes "correlation at one time point in the same cluster between two different subjects:"

$$\rho = \frac{\sigma_c^2 + \sigma_{ct}^2}{\sigma_c^2 + \sigma_{ct}^2 + \sigma_s^2 + \sigma_{st}^2} \qquad \text{(Teerenstra et al., 2012, Equation 6)}$$

The definition for r relates to cluster means. To arrive at the correlation for an individual in the cluster, consider n=1. Thus, the correlation between an individual's paired observations at baseline and follow up, is equal to $Corr(y_{i1}, y_{i2}) = \rho_c \rho + \rho_s(1-\rho)$.

Therefore, the off-diagonal covariances in A, $Cov(y_{i1}, y_{i2})$, are equal to $Corr(y_{i1}, y_{i2}) * \sigma_y^2 = \rho_c \rho \sigma_y^2 + \rho_s(1-\rho)\sigma_y^2$

and the elements of A are:

$$A = \begin{matrix} Var(y) & Cov(y_{i1}, y_{i2}) \\ Cov(y_{i1}, y_{i2}) & Var(y) \end{matrix} = \begin{matrix} \sigma_y^2 & \rho_c \rho \sigma_y^2 + \rho_s(1-\rho)\sigma_y^2 \\ \rho_c \rho \sigma_y^2 + \rho_s(1-\rho)\sigma_y^2 & \sigma_y^2 \end{matrix}$$

The submatrices B are identical 2x2 submatrices that describe covariances between observations obtained on pairs of different individuals (i≠i`) within the same cluster. The submatrices B have the form:

$$B = \begin{matrix} Cov(y_{i1}, y_{i'1}) & Cov(y_{i1}, y_{i'2}) \\ Cov(y_{i2}, y_{i'1}) & Cov(y_{i2}, y_{i'2}) \end{matrix} = \begin{matrix} \rho\sigma_y^2 & \rho_c\rho\sigma_y^2 \\ \rho_c\rho\sigma_y^2 & \rho\sigma_y^2 \end{matrix} = \begin{matrix} \sigma_c^2 + \sigma_{ct}^2 & \sigma_c^2 \\ \sigma_c^2 & \sigma_c^2 + \sigma_{ct}^2 \end{matrix}$$

These results follow from applying the definitions of ρ (Tereenstra, et al.,2012, Equation 6)

$$\rho\sigma_y^2 = \frac{\sigma_c^2 + \sigma_{ct}^2}{\sigma_c^2 + \sigma_{ct}^2 + \sigma_s^2 + \sigma_{st}^2} * [\sigma_c^2 + \sigma_{ct}^2 + \sigma_s^2 + \sigma_{st}^2] = \sigma_c^2 + \sigma_{ct}^2$$

and of $\rho_c$ (Tereenstra, et al., 2012, Equation 2)

$$\rho_c \rho \sigma_y^2 = \frac{\sigma_c^2}{\sigma_c^2 + \sigma_{ct}^2}(\sigma_c^2 + \sigma_{ct}^2) = \sigma_c^2.$$

Having arrived at these expected values for elements of the V matrix, the analyst can later verify that the GLMM faithfully produces them when applied to an exemplary dataset that is appropriate for this design. Before we demonstrate how to estimate sample size and power in a GLMM, we will demonstrate how to calculate and apply an appropriate design effect.

### *Applying a design effect to estimate sample size*

The design effect that Teerenstra et al. (2012) derive for the pre-post design (see also Hooper et al., 2016, Equation 10) applies again to this example:

$$DE = [1 + (n-1)\rho] * (1 - r^2)$$

Where $r = \frac{n\rho}{1+(n-1)\rho}\rho_c + \frac{1-\rho}{1+(n-1)\rho}\rho_s$ and where cluster autocorrelation ($\rho_c$) and subject autocorrelation ($\rho_s$) are as previously defined.

Previous examples have incorporated assumptions of total variance of 25, a cluster size of 10, an ICC ($\rho$) of 0.1 and cluster autocorrelation ($\rho_c$) of 0.4. Adding an assumption for the value of $\rho_s$ (0.6), then the DE for a cohort CRT where the outcome is measured twice on each individual, is:

$[1 + (n - 1)\rho] = 1.9$ (as previously)

$$r = \frac{n\rho}{1 + (n - 1)\rho}\rho_c + \frac{1 - \rho}{1 + (n - 1)\rho}\rho_s = \frac{(10)(0.1)}{1 + (9)(0.1)}(0.4) + \frac{1 - 0.1}{1 + (9)(0.1)}(0.6)$$

$$r = \frac{0.40 + 0.54}{1.9} = 0.495$$

$$DE = [1 + (n - 1)\rho] * (1 - r^2) = (1.9)(1-0.495^2) = 1.435$$

Multiplying the total sample size for the RCT, 128, by the design effect of 1.435 yields a sample estimate for the CRT of 183.67. We take this number to represent the total number of observations. If each cluster measures the outcome on 10 individuals at two time points (20 observations per cluster), then nine provider clusters (five randomly assigned to provide standard care, and four assigned to provide the novel treatment) will generate 180 observations. This design should afford power that nearly matches the value of 0.789 predicted by PROC GLMPOWER for an RCT with 128 independent observations.

## *Reproducing the results in a GLMM*

Table 2 contains the DATA step that generates an appropriate exemplary dataset for this cohort CRT. The syntax resembles that which generated data for the cross-sectional CRT in Example 4. The sequence of DO loops ("times within individual" versus "individuals within times") does not affect the mean structure.

To construct a GLMM that accommodates the cohort CRT's multilevel design (two measurement on each individual, and 10 individuals nested within each cluster), we must also specify expected values for the four variance components: $\sigma_c^2, \sigma_{ct}^2, \sigma_{st}^2, \sigma_s^2$.

Starting with Equation (6) in Teerenstra et al. (2012):

$$\rho = \frac{\sigma_c^2 + \sigma_{ct}^2}{\sigma_c^2 + \sigma_{ct}^2 + \sigma_s^2 + \sigma_{st}^2} = \frac{\sigma_c^2 + \sigma_{ct}^2}{\sigma_y^2}, \text{ therefore } \sigma_c^2 + \sigma_{ct}^2 = \rho\sigma_y^2$$

It follows that:
$$1 - \rho = \frac{\sigma_c^2 + \sigma_{ct}^2 + \sigma_s^2 + \sigma_{st}^2}{\sigma_c^2 + \sigma_{ct}^2 + \sigma_s^2 + \sigma_{st}^2} - \frac{\sigma_c^2 + \sigma_{ct}^2}{\sigma_c^2 + \sigma_{ct}^2 + \sigma_s^2 + \sigma_{st}^2} = \frac{\sigma_s^2 + \sigma_{st}^2}{\sigma_y^2}, \text{ therefore } \sigma_s^2 + \sigma_{st}^2 = (1 - \rho)\sigma_y^2$$

We then use Equation 2 from Teerenstra et al. (2012) to express the variance component $\sigma_{ct}^2$ in terms of $\sigma_c^2$ and the component $\sigma_{st}^2$ int terms of $\sigma_s^2$:

$$\rho_c = \frac{\sigma_c^2}{\sigma_c^2 + \sigma_{ct}^2}, \text{ therefore } \sigma_{ct}^2 = \frac{(1-\rho_c)}{\rho_c}\sigma_c^2$$

Because $\sigma_c^2 + \sigma_{ct}^2 = \rho\sigma_y^2$, then $\sigma_c^2 + \frac{(1-\rho_c)}{\rho_c}\sigma_c^2 = \rho\sigma_y^2$, and $\sigma_c^2 = \rho_c\rho\sigma_y^2$

$$\rho_s = \frac{\sigma_s^2}{\sigma_s^2 + \sigma_{st}^2}, \text{ therefore } \sigma_{st}^2 = \frac{(1-\rho_s)}{\rho_s}\sigma_s^2$$

Because $\sigma_s^2 + \sigma_{st}^2 = (1-\rho)\sigma_y^2$, then $\sigma_s^2 + \frac{(1-\rho_s)}{\rho_s}\sigma_s^2 = (1-\rho)\sigma_y^2$, and $\sigma_s^2 = \rho_s(1-\rho)\sigma_y^2$

By substituting into these equivalences the expected values for $\rho$, $\rho_c$, $\rho_s$ and $\sigma_y^2$, we generate values for all four variance components. This example assumes a value for $\rho$, the ICC, of 0.1; a total variance $\sigma_y^2 = 25$; cluster autocorrelation $\rho_c = 0.4$; and subject autocorrelation $\rho_s = 0.6$. Then:

$$\sigma_c^2 = \rho_c\rho\sigma_y^2 = (0.4)(0.1)(25) = 1$$
$$\sigma_c^2 + \sigma_{ct}^2 = \rho\sigma_y^2, \text{ and so } \sigma_{ct}^2 = \rho\sigma_y^2 - \sigma_c^2 = (0.1)(25) - 1 = 1.5$$
$$\sigma_s^2 = \rho_s(1-\rho)\sigma_y^2 = (0.6)(0.9)(25) = 13.5$$
$$\sigma_s^2 + \sigma_{st}^2 = (1-\rho)\sigma_y^2, \text{ and so } \sigma_{st}^2 = (1-\rho)\sigma_y^2 - \sigma_s^2 = (0.9)(25) - 13.5 = 9$$

Table 1 shows the PROC MIXED step that generates the appropriate GLMM. The procedure incorporates three RANDOM statements that define random intercepts whose variances are $\sigma_c^2$, $\sigma_{ct}^2$ and $\sigma_s^2$. The PARMS statement specifies values for these "covariance parameters" in the order the RANDOM statements appear. No separate RANDOM statement defines the fourth parameter $\sigma_{st}^2$. The PARMS statement assigns a value to this fourth variance component so that the four components sum to $\sigma_y^2$. The procedure regards and reports $\sigma_{st}^2$ as a residual variance.

To verify that the syntax for the RANDOM and PARMS statements produce the expected variance structure, investigators must check the V and VCORR matrices produced by PROC GLIMMIX or PROC MIXED. Depicted below are the first 10 rows and 10 columns of the cluster-specific V matrix induced for this example. A full V matrix for this example has 20 rows and 20 columns; variances and covariances are depicted only for responses among five of a cluster's ten individuals. The submatrices previously labelled as A are shaded and the 2x2 submatrices previously labelled as B are located off this diagonal "backbone."

| 25.0 | 14.5 | 2.5 | 1.0 | 2.5 | 1.0 | 2.5 | 1.0 | 2.5 | 1.0 |
| 14.5 | 25.0 | 1.0 | 2.5 | 1.0 | 2.5 | 1.0 | 2.5 | 1.0 | 2.5 |
| 2.5 | 1.0 | 25.0 | 14.5 | 2.5 | 1.0 | 2.5 | 1.0 | 2.5 | 1.0 |
| 1.0 | 2.5 | 14.5 | 25.0 | 1.0 | 2.5 | 1.0 | 2.5 | 1.0 | 2.5 |
| 2.5 | 1.0 | 2.5 | 1.0 | 25.0 | 14.5 | 2.5 | 1.0 | 2.5 | 1.0 |

| 1.0 | 2.5 | 1.0 | 2.5 | 14.5 | 25.0 | 1.0 | 2.5 | 1.0 | 2.5 |
|---|---|---|---|---|---|---|---|---|---|
| 2.5 | 1.0 | 2.5 | 1.0 | 2.5 | 1.0 | 25.0 | 14.5 | 2.5 | 1.0 |
| 1.0 | 2.5 | 1.0 | 2.5 | 1.0 | 2.5 | 14.5 | 25.0 | 1.0 | 2.5 |
| 2.5 | 1.0 | 2.5 | 1.0 | 2.5 | 1.0 | 2.5 | 1.0 | 25.0 | 14.5 |
| 1.0 | 2.5 | 1.0 | 2.5 | 1.0 | 2.5 | 1.0 | 2.5 | 14.5 | 25.0 |

The elements of V reflect the expected values. These include the covariance between observations obtained on the same individuals before and after intervention: $Cov(y_{i1}, y_{i2}) = \rho_c \rho \sigma_y^2 + \rho_s(1-\rho)\sigma_y^2 = [0.04 + 0.54]25 = 14.5$.

## *Comparing results from the two approaches*

Application of the design effect predicts that a sample of 183.67 individuals will afford power of roughly 0.789 (the estimate generated using PROC GLMPOWER) to detect the hypothesized the hypothesized arm*time interaction. Table 1 shows that the GLMM estimates the power of this cohort CRT to detect the hypothesized interaction to be 0.830. This estimate is larger than the one predicted by application of the design effect, even though the design involves only 180 individuals. Among the examples considered thus far, this one reveals the largest difference in the two approaches' power estimates.

## 6. A "cross sectional" SWD where the outcome is measured repeatedly at each "step," some before and some after intervention, in clusters that recruit independent groups at each step.

The stepped wedge design is a type of crossover design where crossover is unidirectional, generally from a control or baseline phase to a treatment phase. Individuals in an SWD are generally members of natural clusters, and the sequence of crossover is randomized according to those clusters. For example, in a design with six clusters, individuals in all clusters are measured initially in the control phase (T=0). Three clusters are randomly assigned to initiate the intervention at the first step; their members are measured after initiation of the intervention at time T=1, and again at T=2. The other three clusters, which were randomly assigned to await the transition to the intervention until the second step, are measured before the transition at T=1 and again after initiating the intervention, at time T=2.

## *Statistical model, mean and variance structure*

Hussey and Hughes (2007) propose the following statistical model for the mean response of cluster i at time j in a SWT:

$$\mu_{ij} = \mu + \alpha_c + \beta_t + X_{ij}\theta$$

Cluster assignment (c) is random, so that $\alpha_c$ is a cluster-specific random intercept. Times of measurement (t) are fixed. $X_{ij}$ is an indicator variable that denotes the phase of treatment (0=control; 1=intervention). $\theta$ is the estimated treatment effect, which is also assumed to be fixed.

To make the model's notation consistent with that of previous examples, we rewrite it, replacing the coefficients that estimate fixed effects, $\theta$ and $\beta_t$, with $\beta_1$ and $\beta_2$. The random intercept $\alpha_c$ is replaced by $b_c$. The model, in the revised notation, is:

$$y_{ict} = \beta_0 + \beta_1 X_{ct} + \beta_2 \text{time} + b_c + e_{ict}$$

Like those displayed for earlier examples, this model identifies sources of variability in the value of the outcome y for subject i in clinic or cluster c at time t under intervention $X_{ct}$. Coefficients associated with fixed effects are denoted by β and those associated with random effects by b.

The model contains a fixed effect for time. [Hussey and Hughes](#) (2007) point out that, because secular trends could confound the effect of an intervention that is initiated at different times for different clusters, a GLMM used to analyze an SWD should include time as a fixed effect. [Woertman et al](#). (2013) reinforce the fact Hussey and Hughes' model accounts or adjust for "external time trends."

[Woertman et al.](#) (2013) also point out that the model makes appropriate adjustments only if secular effects, if they exist, apply equally to all clusters. The model includes no term to assess cluster-time interactions. The model is valid only if the outcome is unaffected by the time at which a cluster crosses over. Clusters that transition earlier cannot differ in their mean response from clusters that transition later. The model also assumes that any secular effect on a cluster's mean response is the same for every cluster, regardless of the intervention they experience, or the group of individuals measured at a given time.

The model treats clusters, but not individuals within clusters, as a random effect. Thus, the model is suited to an SWD in which responses are correlated by virtue of cluster membership, and where each set of measurements within a cluster is obtained from a unique and independent group of individuals at each time point. This model does not accommodate within-subject correlation over times of measurement. Example 7 describes a longitudinal or "cohort" SWD in which the same groups of individuals are followed over time within each cluster.

The variance of the response y for subject i in cluster c at time t is:

$$\text{Var}(y_{ict}) = \text{Var}(\beta_0 + \beta_1 X_{ct} + \beta_2 \text{time} + b_c + e_{ict}) = \sigma_c^2 + \sigma_e^2$$

Thus, the variance has two components, the between-cluster variance, $\text{Var}(b_c) = \sigma_c^2$ and the within-cluster or error variance, $Var(e_{ict}) = \sigma_e^2$. As in previous examples, the intraclass correlation coefficient ICC is $\rho = \frac{\sigma_c^2}{\sigma_c^2 + \sigma_e^2}$.

The [Hussey and Hughes](#) (2007) model assumes that cluster autocorrelation is equal to 1. To accommodate an expected value for cluster association ($\rho_c$) that is less than one, the model for the cross-sectional CRT is revised as:

$$y_{ict} = \beta_0 + \beta_1 X_{ct} + \beta_2 time + b_c + b_{ct} + e_{ict}$$

where values for the two random intercepts define $\rho_c = \frac{Var(b_c)}{Var(b_c) + Var(b_{ct})} = \frac{\sigma_c^2}{\sigma_c^2 + \sigma_{ct}^2}$

### *Applying a design effect to estimate sample size*

[Woertman et al.](#) (2013) propose a design effect for a stepped wedge trial (DE$_{SW}$) wherein each clinic cluster observes a different set of individuals at each time of measurement.

$$\text{DE}_{SW} = \frac{1 + \rho(ktn + bn - 1)}{1 + \rho(\frac{1}{2}ktn + bn - 1)} * \frac{3(1-\rho)}{2t(k - \frac{1}{k})}$$

where k is the number of steps in the stepped wedge, b is the number of baseline (pre-intervention) measurements obtained on each cluster; t is the number of post-intervention measurements obtained on each cluster; n is the number of individuals per cluster; and ρ denotes the intracluster correlation coefficient (ICC). [Baio](#) et al. (2015) caution that this DE$_{SW}$ must be further adjusted to obtain "the overall sample size in terms of participants (each contributing one measurement)."

To illustrate the use of the DE$_{SW}$, we consider a stepped wedge design with 2 steps (k=2), single measurements at each time point (b=1 and t=1), and a cluster size of 5. As with the previous examples, we will assume a realistic and conservative ICC of 0.10.

Inserting these values into the formula for DE$_{SW}$:

$$\text{DE}_{SW} = \frac{1 + \rho(ktn + bn - 1)}{1 + \rho(\frac{1}{2}ktn + bn - 1)} * \frac{3(1-\rho)}{2t(k - \frac{1}{k})} = \frac{(1 + (.1)(10 + 5 - 1)) * 3(0.9)}{(1 + (0.1)(5 + 5 - 1)) * 2(2 - \frac{1}{2})}$$

$$= \frac{6.48}{5.7} = 1.137$$

Example 1 considered a two-arm RCT where the outcome is measured once, following a period of intervention. It required participation of 34 people, divided between two groups of 17, to afford power of 0.807 to detect a hypothesized between-group difference of 5 units, assuming a common standard deviation of 5 units. Following the counsel of Baio et al. (2015), we multiply the $DE_{SW}$ of 1.137 by (b+kt)=3, and use this product to multiply the total sample size for the RCT. The resulting sample size estimate for the SWD 1.137*3*34 or 116. This represents the total number of individuals measured across all clusters and across all times of measurement. A design that involves eight clinics (clusters), each of which measures the outcome on five individuals at each of the three time points (15 observations per clinic), will involve 120 individuals. If four of the eight clinics initiate the intervention at each of the k=2 steps, then the study should afford slightly more power than that calculated for the RCT with 34 independent observations.

Hooper et al. (2016, p. 2722) regard the SWD to be a member of a family of "longitudinal cluster randomised trials." They provide a general formula to obtain DE for studies within this family, and they identify the $DE_{SW}$ of Woertman et al. (2013) as applying to the special case where cluster autocorrelation $\rho_c$ is equal to 1, and subject autocorrelation $\rho_s$ is equal to zero. They also cite an approach to obtain a DE for the special case of an SWD like this example, where the outcome is measured three times: once at baseline (b=1) and once (t=1) after each of the k=2 steps. For this type of SWD, de Hoop et al. (2015, Equation 12, pp.37-38) state that one can multiply the familiar design effect due to clustering [1+(n-1)ρ] by the variance inflation factor $[1 - \frac{2r^2}{1+r}]$, where r is the familiar quantify for correlation formulated by Teerenstra et al. (2012), $r = \frac{n\rho}{1+(n-1)\rho}\rho_c + \frac{1-\rho}{1+(n-1)\rho}\rho_s$. For this special case, If $\rho_c$=1 and $\rho_s$=0 (because the design is cross-sectional), de Hoop's formula yields the same design effect of 1.137 as Woertman's.

## *Reproducing the results in a GLMM*

Table 2 contains a DATA step that generates an exemplary dataset whose mean structure demonstrates an intervention effect of 5 units. The data set is of the size specified by application of the design effect of Woertman et al. (2013) and de Hoop et al. (2015). Table 1 shows syntax, constructed using SAS PROC GLIMMIX, that creates a GLMM tailored to the Hussey and Hughes (2007) model. It includes a fixed effect for time that accounts and adjusts for the potential that external or secular "time trends" could confound the association between an intervention and an outcome. As described earlier, the model and the $DE_{SW}$ created to accompany it assume there is no cluster-time interaction. Secular effects, if they exist, apply equally to all clusters.

The exemplary data set specifies the structure of expected mean responses. The investigator must additionally state the expected values for the variances $\sigma_c^2$ and $\sigma_e^2$. Maintaining the assumptions that the ICC is 0.1 and the total variance $\sigma_y^2$ is 25, the expected values are, respectively, 2.5 and 22.5. The expected cluster-specific V matrix has 15 rows and columns. The structure of V is similar to the one described for the cross-sectional CRT (Example 4) and is composed of submatrices A and B.

$$V = \begin{matrix} A & B & B \\ B & A & B \\ B & B & A \end{matrix}$$

The submatrices A, like those described for the cross-sectional RCT, have five rows and five columns, one for each individual within the cluster.

| $\sigma_c^2 + \sigma_e^2$ | $\sigma_c^2$ | $\sigma_c^2$ | $\sigma_c^2$ | $\sigma_c^2$ |
|---|---|---|---|---|
| $\sigma_c^2$ | $\sigma_c^2 + \sigma_e^2$ | $\sigma_c^2$ | $\sigma_c^2$ | $\sigma_c^2$ |
| $\sigma_c^2$ | $\sigma_c^2$ | $\sigma_c^2 + \sigma_e^2$ | $\sigma_c^2$ | $\sigma_c^2$ |
| $\sigma_c^2$ | $\sigma_c^2$ | $\sigma_c^2$ | $\sigma_c^2 + \sigma_e^2$ | $\sigma_c^2$ |
| $\sigma_c^2$ | $\sigma_c^2$ | $\sigma_c^2$ | $\sigma_c^2$ | $\sigma_c^2 + \sigma_e^2$ |

The diagonal members of A are all equal to the specified total variance of 25. The off-diagonal covariances are all equal to 2.5, reflecting the specified ICC of 0.1.

The elements of the 5x5 submatrices B, because the cluster autocorrelation is assumed to be $\rho_c=1$, are all covariances equal to $\rho_c \rho \sigma_y^2 = 2.5$.

To generate the appropriate variance structure, the PROC GLIMMIX step (Table 1) incorporates a RANDOM statement that defines the single random intercept $b_c$ for each cluster, and a PARMS statement that specifies values for the variance components $\sigma_c^2$ and $\sigma_e^2$.

## *Comparing results from the two approaches*

Application of the design effect predicts that a total sample of 116 individuals, measured across all clusters and across all times of measurement, will match the power of 0.807 calculated by PROC POWER to detect the hypothesized between-arm difference of 5 units in an RCT with a total of 34 independent observations. Applied to a cross-sectional SWD with 120 individuals (8 clinics, each with 5 individuals, with measurements at three points in time), the GLMM estimates the design's power at 0.836. The estimates' similarity is evidence that the approaches validate one another.

The $DE_{SW}$ proposed by Woertman, et al. (2013) applies only to situations "where the same number of clusters switches at each step and where the number of measurements after each step is constant as well (p.754)." It also assumes a cluster autocorrelation of 1; correlations between measurements on individuals in the same cluster are assumed to be the same even in cross-sections observed at different points in time. The DE proposed by de Hoop (2015) permits different assumptions concerning cluster autocorrelation. However, it applies only to situations where cross-sections are measured three times over two steps.

The direct estimation of sample size and power in a GLMM has an advantage in that it can accommodate designs wherein practical considerations might necessitate, for example, switching of different numbers of clusters at different steps. An exemplary dataset and a complementary

GLMM could also be constructed to accommodate cluster*time interactions if one has reason to believe that individuals in clusters respond differently to the intervention depending on the secular time at which the clusters cross over.

## 7. A "cohort" SWD where the outcome is measuredly repeatedly at each "step," some before and some after intervention, in clusters that follow and observe the same individuals over time.

The preceding example of an SWD involves what **Baio** et al. (2015) call a "repeated cross sectional design" in which "measurements are obtained at discrete times from different individuals." This contrasts with a "cohort SWD" in which clusters follow intact groups or cohorts of individuals over the course of repeated measurements. These designs are potentially more powerful because each individual serves as his or her own control.

*Statistical model, mean and variance structure*

We modify the statistical model for the cross-sectional SWT that was inspired by Hussey and Hughes (2007) so that it accommodates correlations both at the level of the cluster (cluster autocorrelation) and within individuals who are measured repeatedly over time (subject autocorrelation):

$$y_{ict} = \beta_0 + \beta_1 X_{ct} + \beta_2 time + b_c + b_{ct} + b_s + b_{st}$$

This model includes the same four random intercepts defined for the cohort CRT (example 5). The random effects imply four variance components.

$$\begin{aligned}\text{Var}(y_{ict}) &= \text{Var}(\beta_0 + \beta_1 X_{ct} + \beta_2 time + b_c + b_{ct} + b_s + b_{st}) = \sigma_y^2 \\ &= \sigma_c^2 + \sigma_{ct}^2 + \sigma_{st}^2 + \sigma_s^2\end{aligned}$$

The same definitions introduced in the discussion of the cohort CRT apply to this cohort SWD: an intraclass correlation coefficient $\rho$, and coefficients for cluster autocorrelation ($\rho_c$) and for subject autocorrelation ($\rho_s$). The cluster-specific V matrix, which describes variances and covariances among the repeated observations on the cluster's n individuals, is composed of submatrices A and B. For a cluster with n=3 individuals, V is of the block diagonal form:

$$\begin{matrix} A & B & B \\ B & A & B \\ B & B & A \end{matrix}$$

The number of submatrices A is equal to the number of individuals in the cluster. Each submatrix A has b+kt rows and columns, which list the expected variances and covariances among the b+kt repeated measurements obtained on a single individual. For a cohort SWD with k=2 steps, and where each individual is measured once before intervention (b=1) and once after each step (t=1), the A submatrices are 3x3 and of the form described in detail for the cohort CRT:

$$A = \begin{matrix} & & V & \\ & Var(y) & Cov(y_{i1}, y_{i2}) & Cov(y_{i1}, y_{i3}) \\ & Cov(y_{i2}, y_{i1}) & Var(y) & Cov(y_{i2}, y_{i3}) \\ & Cov(y_{i3}, y_{i1}) & Cov(y_{i3}, y_{i2}) & Var(y) \end{matrix}$$

$$= \begin{matrix} \sigma_y^2 & \rho_c\rho\sigma_y^2 + \rho_s(1-\rho)\sigma_y^2 & \rho_c\rho\sigma_y^2 + \rho_s(1-\rho)\sigma_y^2 \\ \rho_c\rho\sigma_y^2 + \rho_s(1-\rho)\sigma_y^2 & \sigma_y^2 & \rho_c\rho\sigma_y^2 + \rho_s(1-\rho)\sigma_y^2 \\ \rho_c\rho\sigma_y^2 + \rho_s(1-\rho)\sigma_y^2 & \rho_c\rho\sigma_y^2 + \rho_s(1-\rho)\sigma_y^2 & \sigma_y^2 \end{matrix}$$

Like the cohort CRT (example 5), the B submatrices are of the form:

$$B = \begin{matrix} \rho\sigma_y^2 & \rho_c\rho\sigma_y^2 & \rho_c\rho\sigma_y^2 \\ \rho_c\rho\sigma_y^2 & \rho\sigma_y^2 & \rho_c\rho\sigma_y^2 \\ \rho_c\rho\sigma_y^2 & \rho_c\rho\sigma_y^2 & \rho\sigma_y^2 \end{matrix} = \begin{matrix} \sigma_c^2 + \sigma_{ct}^2 & \sigma_c^2 & \sigma_c^2 \\ \sigma_c^2 & \sigma_c^2 + \sigma_{ct}^2 & \sigma_c^2 \\ \sigma_c^2 & \sigma_c^2 & \sigma_c^2 + \sigma_{ct}^2 \end{matrix}$$

### *Applying a design effect to estimate sample size*

The design effect proposed by Woertmann et al. (2013) for the cross-sectional SWD requires a variety of assumptions that make it inapplicable to the cohort SWD. Most important, it assumes that that each clinic cluster observes a unique and independent group of individuals at each step. While it recognizes that measurements are correlated by virtue of common membership within a cluster, Woertmann's $DE_{SW}$ cannot account for within-subject correlation (or subject autocorrelation) across times of measurement.

In fact, Hemming et al. (2015) comment that "as yet, there is no specific adaptation of design effects [nor implementation in a statistical package] for calculating the power or sample size in a cohort stepped wedge trial." Likewise, Baio (2015) states that "reliable sample size algorithms for more complex designs, such as those using cohorts rather than cross-sectional data, have not yet been established."

However, the DE proposed by de Hoop et al. (2015) applies to this example, if only because this SWD fits a special case they describe, one that involves exactly three measurements, a single baseline measurement (b=1) and single follow-up measures (t=1) obtained at k=2 steps. The DE multiplies the familiar design effect due to clustering randomizing [1+(n-1)ρ] by the design effect due to repeated assessment $[1 - \frac{2r^2}{1+r}]$, where r is the familiar quantity for correlation formulated by Teerenstra et al. (2012).

$$DE = [1 + (n-1)\rho] * (1 - \frac{2r^2}{1+r})$$

$$r = \frac{n\rho}{1+(n-1)\rho}\rho_c + \frac{1-\rho}{1+(n-1)\rho}\rho_s$$

To facilitate comparison, we maintain assumptions of a cluster size of 5, total variance of 25, an ICC ($\rho$) of 0.1 cluster autocorrelation ($\rho_c$) of 0.4 and subject autocorrelation ($\rho_s$) of 0.6. Then, for this cohort SWD with three measurements:

$$r = \frac{n\rho}{1+(n-1)\rho}\rho_c + \frac{1-\rho}{1+(n-1)\rho}\rho_s = \frac{(5)(0.1)}{1+(4)(0.1)}(0.4) + \frac{1-0.1}{1+(4)(0.1)}(0.6) = 0.529$$

$$[1+(n-1)\rho] = 1.4$$

$$DE = [1.4] * \left(1 - \frac{2(0.529)^2}{1+0.529}\right) = 0.888$$

Multiplying the total sample size for the RCT, 34, by the design effect of 0.888, yields a sample estimate for the cohort SWD of 30.2. This number represent the total number of individuals. Because each individual is observed three times, the required sample size involves a total of 90.6 observations. A cohort SWD with six clusters, with 5 individuals in each cluster, should afford power that nearly matches the value of 0.807 predicted by PROC GLMPOWER for an RCT with 34 independent observations.

### *Estimating power directly in a GLMM*

Table 2 depicts a SAS DATA step that creates exemplary dataset with the hypothesized mean structure. The DATA step closely resembles the one used to create the exemplary dataset for the cross-sectional SWD discussed in example 6. They differ in that, in this example, the first of two nested DO sequences defines individuals, and the second simulates three repeated observations for each individual. To create the exemplary dataset for the cross-sectional SWD, the nested DO sequences first define times of measurements, then simulate observations for individuals at each time of measurement.

This example involves 6 clusters, each with 5 individuals. Three clusters switch from the standard to the novel intervention at the first step, and three switch at the second step. Because the outcome is measured three times on each individual, the design involves 30 individuals and 90 observations.

Table 2 depicts the PROC MIXED syntax that generated the appropriate GLMM for the cohort SWD. We rely on the formulae of Teerenstra et al. (2012), as we did for the cohort CRT (example 5), to arrive at the expected values for the variance components that reflect the assumed values for cluster and subject autocorrelation.

The PROC MIXED step incorporates three RANDOM statements to define random intercepts, whose respective variance components are equal to the expected values for $\sigma_c^2$, $\sigma_{ct}^2$ and $\sigma_s^2$. The PARMS statement specifies these values in the order that the RANDOM statements appear. No separate RANDOM statement defines the fourth parameter $\sigma_{st}^2$. The PARMS statement assigns a value to this fourth variance component so that the four components sum to $\sigma_y^2$. PROC MIXED regards $\sigma_{st}^2$ as a residual variance. Because the design involves two levels, individuals within clusters, the MODEL statement includes the option DDFM=BETWITHIN. This causes the denominator degrees of freedom, which are used to estimate F statistics, to be determined using the "between-within" method that is recommended for a multilevel design.

PROC MIXED induces a cluster-specific V matrix of the expected dimensions, with 15 rows and columns. A 9x9 portion of the V matrix, which lists variances and covariances for three individuals, appears below. The 3x3 submatrices A, which are shaded and arrayed on the main diagonal, reflect repeated observations on individuals. The 3x3 submatrices B, which are off the main diagonal, illustrate covariances among observations obtained on three occasions but on different individuals within the cluster. The values of the matrix' elements reflect those assumed for ρ, $ρ_c$ and $ρ_s$.

| 25.0 | 14.5 | 14.5 | 2.5  | 1.0  | 1.0  | 2.5  | 1.0  | 1.0  |
|------|------|------|------|------|------|------|------|------|
| 14.5 | 25.0 | 14.5 | 1.0  | 2.5  | 1.0  | 1.0  | 2.5  | 1.0  |
| 14.5 | 14.5 | 25.0 | 1.0  | 1.0  | 2.5  | 1.0  | 1.0  | 2.5  |
| 2.5  | 1.0  | 1.0  | 25.0 | 14.5 | 14.5 | 2.5  | 1.0  | 1.0  |
| 1.0  | 2.5  | 1.0  | 14.5 | 25.0 | 14.5 | 1.0  | 2.5  | 1.0  |
| 1.0  | 1.0  | 2.5  | 14.5 | 14.5 | 25.0 | 1.0  | 1.0  | 2.5  |
| 2.5  | 1.0  | 1.0  | 2.5  | 1.0  | 1.0  | 25.0 | 14.5 | 14.5 |
| 1.0  | 2.5  | 1.0  | 1.0  | 2.5  | 1.0  | 14.5 | 25.0 | 14.5 |
| 1.0  | 1.0  | 2.5  | 1.0  | 1.0  | 2.5  | 14.5 | 14.5 | 25.0 |

*Power predictions from the GLMM*

Assuming the same values for ρ (0.1), $ρ_c$ (0.4) and $ρ_s$ (0.6) used in previous examples, the design affords power of 0.819 to detect the hypothesized intervention effect of 5 units. The effort expended to follow a cohort of individuals over time has a benefit; the cohort SWD affords power similar to that afforded by a cross-sectional SWD. However, the cohort SWD requires recruitment of fewer clusters.

The GLMM's estimate for power of 0.819 compares with the power estimate of 0.807 for the RCT whose sample size was inflated using the design effect proposed by de Hoop et al. (2015). We note that this DE applies to this example because this cohort SWD features k=2 steps and three occasions of measurement. The DE also accords with another restriction, which is that the number of clusters that switch at either of the two steps must be equal. Thus, the DE proposed by de Hoop applies to a limited range of cohort SWD. On the other hand, the DE proposed by Woertman et al. (2013) cannot apply to a cohort SWD because it does not accommodate subject autocorrelation.

The limitations of DE available for cohort SWD make the use of GLMM attract for direct prediction of power. GLMM can be applied to exemplary datasets that involve multiple steps, multiple measurements within steps, differences in the number of clusters that switch at each step, and differences in the number of individuals per cluster.

## Conclusion

This tutorial has demonstrated and compared results from two approaches that estimate sample size and power for studies with correlated and clustered observations. By producing similar estimates of power, the two approaches generally validate one another. When they differ, estimates of power from the GLMM are slightly larger than those obtained by application of the design effect. This finding may accord with a comment made by Hooper et al. (2016, p. 4726), who remarked that "asymptotic sample size formulae will underestimate required sample size when the number of clusters is small."

While Hooper et al. (2016) have made progress in formulating a general approach to defining design effects for studies with clustered data, readily applicable expressions for DE have not been published for all designs. For designs for which no DE is available, a GLMM can be constructed to assess sample size and directly estimate power. The GLMM has additional advantages in that investigators can assess sample size and power using the same statistical model they will apply to the study's data after they are collected. Moreover, GLMM can accommodate variance structures that are more complex than the ones illustrated in this tutorial. GLMM can also accommodate design features that arise in real world settings, including unequal cluster sizes, different numbers of clusters assigned to different treatments in the arms of a CRT, or assigned to switch treatments at steps in the SWD.

This tutorial has focused on applying design effects or constructing GLMM to address an outcome that is plausibly normally distributed. However, the two approaches apply to outcomes that follow other distributions. Calculating $\chi^2$ or t statistics while treating observations as unclustered, then dividing these statistics by the DE or the square root of the DE, respectively, produces appropriate cluster adjusted hypothesis tests (Wears, 2002, p. 333). Similarly, PROC GLIMMIX syntax is readily modified to accommodate outcomes that are measured as proportions, odds, or counts.

# Table 1. Statistical models, SAS procedures and associated power estimates

| Example | Statistical model | PROC statements | Sample size | Power |
|---|---|---|---|---|
| 1. Two-arm RCT, outcome measured once. | $y_{ig} = \beta_0 + \beta_1 arm + e_i$ | ```proc power;   twosamplemeans     test=diff sides=2 alpha=0.05     groupmeans= 59\| 54     stddev=5     npergroup=17     power=.; run;``` | Ntotal=34 | 0.807 |
| | | ```proc glimmix data=one noprofile; class treatment; model mean=treatment/solution; parms (25) / hold=1; ods output tests3=fstats; run;``` | Ntotal=34 | 0.807 |
| 2. Two-arm arm CRT, outcome measured once | $y_{icg} = \beta_0 + \beta_1 arm + b_c + e_i$ | Application of DE | DE=1.5 Ntotal=1.5*34=51 | |
| | | ```proc glimmix data=one noprofile; class treatment clinicid patientid; model mean=treatment / solution; random intercept / sub=clinicid v vcorr; parms (2.5) (22.5) / hold=1,2; ods output tests3=fstats; run;``` | Ntotal=54 (6 indiv. in each of 9 clusters) | 0.831 |
| | | | Ntotal=48 (6 indiv. in each of 8 clusters) | 0.788 |
| | | | Ntotal=51 (6 or 7 indiv. in each of 8 clusters) | 0.803 |
| 3. Two arm RCT, outcome measured twice in independent groups | $y_{itg} = \beta_0 + \beta_1 arm + \beta_2 time + \beta_3 (arm * time) + e_i$ | ```data a1;   input treatment time ymean n ;   cards; 1 1 54 32 1 2 56 32 2 1 54 32 2 2 61 32 ; proc glmpower data=a1;   class treatment time ;   model ymean=treatment time treatment*time;   power     stddev = 5     ntotal=124     alpha=.05     power = .; run;``` | Ntotal=128 | 0.789 |
| | | ```proc glimmix data=a2 noprofile; class treatment (ref='1')     time (ref='1'); model mean=treatment time     treatment*time / solution; parms (25) / hold=1; ods output tests3=fstats; run;``` | Ntotal=128 | 0.801 |
| 4. Two-arm "cross-sectional" CRT, outcomes measured twice in independent groups | $y_{ictg} = \beta_0 + \beta_1 arm + \beta_2 time + \beta_3(arm * time) + b_c + b_{ct} + e_i$ | **Application of DE**<br><br>```proc mixed data=c noprofile;
class treatment (ref='1')
     clinic patientid time (ref='1');
model mean=treatment time treatment*time
/ solution ddfm=betwithin;
  random intercept / subject=clinic v
vcorr;
  random intercept / subject=clinic*time
vvcorr;
  parms (1.0) (1.5) (22.5) / hold=1,2,3;
  ods output tests3=fstats v=v
vcorr=vcorr covparms=covparms;
run;``` | DE=1.816<br>Ntotal=128*1.815<br>=232.4<br><br>Ntotal=240 (in 12 clusters with 20 indiv. per cluster) | 0.813 |
| 5. Two-arm "cohort" CRT, outcomes measured twice in groups comprised of the same individuals | $y_{ictg} = \beta_0 + \beta_1 arm + \beta_2 time + \beta_3(arm * time) + b_c + b_{ct} + b_s + b_{st}$ | **Application of DE**<br><br>```proc mixed data=c noprofile;
  class time (ref='1') patientid clinic
treatment (ref='1');
  model mean=treatment time
treatment*time / solution
ddfm=betwithin;
  random intercept / sub=clinic v vcorr;
  random intercept / sub=clinic*time;
  random intercept /
sub=patientid(clinic);
  parms (1) (1.5) (13.5) (9)  / noiter;
  ods output tests3=fstats v=v;
run;``` | DE=1.435<br>Ntotal=128*1.520<br>=183.67<br><br>Ntotal=180 (in 9 clusters with 20 indiv. per cluster) | 0.830 |
| 6. Cross-sectional SWD | $y_{ict} = \beta_0 + \beta_1 X_{ct} + \beta_2 time + b_c + e_{ict}$<br><br>$y_{ict} = \beta_0 + \beta_1 X_{ct} + \beta_2 time + b_c + b_{ct} + e_{ict}$ | **Application of DE**<br><br>```proc glimmix data=xsec_swd order=data
noprofile;
  class step clusterid subjectid time;
  model mean= time intervene  / solution
ddfm=betwithin;
  random intercept / subject=clusterid
v;
  parms (2.5) (22.5) / hold=1,2;
  ods output tests3=fstats;
run;``` | DE=1.137<br>Ntotal=1.137*3*34=116<br><br>Ntotal=120 (in 8 clusters, 5 indiv. per cluster, observed 3 times in independent groups) | 0.836 |
| 7. Cohort SWD | $y_{ict} = \beta_0 + \beta_1 X_{ct} + \beta_2 time + b_c + b_{ct} + b_s + b_{st}$ | **Application of DE**<br><br>```proc mixed data=cohort_swd order=data
noprofile ;
  class step clusterid patientid time;
  model mean= time intervene  / solution
DDFM=betwithin ;``` | DE=0.89<br>Ntotal=0.89*3*34=90.6<br><br>Ntotal=90 (in 6 clusters, 5 | 0.819 |

| | | ```
    random intercept / subject=clusterid
v;           /*specifies sigma^2 c*/
    random intercept /
subject=clusterid*time;
/*specifies sigma^2 ct*/
    random intercept /
subject=patientid(clusterid);
/*specifies sigma^2 s*/
    parms (1) (1.5) (13.5) (9) /
hold=1,2,3,4;
    ods output tests3=fstat v=v;
run;
``` | indiv. per cluster, each indiv. observed 3 times) | |



# Table 2. SAS data steps to create exemplary datasets

| Example | Data step to create exemplary dataset |
|---|---|
| 1. Two-arm RCT, outcome measured once. | ```
DATA one;
array means [2] (59 54);
  do treatment = 1 to 2;
    do subject=1 to 17;
        mean=means[condition];
        output;
    end;
  end;
run;
``` |
| 2. Two-arm arm CRT, outcome measured once | ```
DATA one;
npercluster=6;
array means [2] (59 54);
array clustperarm [2] (5 4);
  do treatment=1 to 2;
    do c=1 to clustperarm[treatment];
    clinicid+1;
       do p=1 to npercluster;
          patientid+1;
         mean=means[treatment];
          output;
        end;
     end;
   end;
run;
``` |
| 3. Two arm RCT, outcome measured twice in independent groups | ```
data a1;
  input treatment time ymean n ;
  cards;
1 1 54 32
1 2 56 32
2 1 54 32
2 2 61 32
;
DATA a2;
  set a1;
  do subject=1 to n;
      mean=ymean;
      output;
  end;
run;
``` |
| 4. Two-arm "cross-sectional" CRT, outcomes measured twice in independent groups | ```
DATA c;
arms=2;
npercluster=10;   /*10 individuals per cluster*/
array means [4]  (54 56 54 61);
array clustperarm [2] (6 6); /*six clinics randomized to either arm of the study*/
  do treatment=1 to arms;
   do c=1 to clustperarm[treatment];
   clinic+1;
      DO TIME=1 TO 2;
        do p=1 to npercluster ;
``` |



| | |
|---|---|
| | ```
          patientid+1;
            mean=means[2*(treatment-1)+time];
              output;
            end;
          end;
        end;
      end;
run;
``` |
| 5. Two-arm "cohort" CRT, outcomes measured twice in groups comprised of the same individuals | ```
DATA c;
arms=2;
npercluster=10;
array means [4]  (54 56 54 61);
array clustperarm [2] (5 4);
  do treatment=1 to arms;
   do c=1 to clustperarm[treatment];
   clinic+1;
     do p=1 to npercluster ;
        patientid+1;
         DO time=1 TO 2;
             mean=means[2*(treatment-1)+time];
             output;
           end;
     end;
   end;
 end;
run;
``` |
| 6. Cross-sectional SWD | ```
data xsec_swd;
  npercluster=5;
  b=1;
  t=1;
  k=2;  /*number of steps*/
  array nk [2]  (4 4);  /*number of clusters that switch at step k*/
  array mm [2] (54 59);  /*expected means before and after step*/
  do step=1 to k;
    do c=1 to nk[step];
      clusterid+1;
      DO TIME=1 TO b+k*t;
        do sub=1 to npercluster;
           subjectid+1;
          if time le step then do;
            intervene=0;
            mean=mm[1];
            output;
           end;
           if time gt step then do;
            intervene=1;
            mean=mm[2];
            output;
           end;
         end;
       end;
    end;
end;
keep step clusterid subjectid time intervene mean ;
run;
``` |

| 7. Cohort SWD | ```
data cohort_swd;
  k=2;  /*number of steps*/
  b=1;     /*number of measurements before first step*/
  t=1;  /*number of measurements after each step*/
  npercluster=5;
  array nk [2]  (3 3);  /*number of clusters that will switch at each of the k steps*/
  array mm [2] (54 59);  /*expected means before and after switch to intervention*/
  do step=1 to k;
    do c=1 to nk[step];
      clusterid+1;
      do sub=1 to npercluster;
      patientid+1;
        DO TIME=1 TO b+k*t;
          if time le step then do;
            intervene=0;
            mean=mm[1];
             output;
           end;
            if time gt step then do;
             intervene=1;
             mean=mm[2];
              output;
            end;
          end;
        end;
     end;
end;
keep step clusterid patientid time intervene mean;
run;
``` |



# Appendix. Power

Power is classically defined as 1-β, the probability of avoiding a Type II error of inference. Power, then, is the probability of correctly rejecting a null hypothesis in favor of an alternative hypothesis that reflects the true state of a population that is the source of the observations one plans to collect in a sample. To estimate power, one must specify the features of an alternative hypothesis.

## *Power defined for the GLMM approach*

The GLMM approach described in this tutorial ([Littell et al.](#) (2006); Stroup ([1999](#), [2013](#)); [Stroup et al. (2018](#), Chapter 14)) relies on the formal definition of power:

$$\text{Power} = \Pr[F\{\text{rank}(L), \text{ddf}, \lambda\} > F\{\text{rank}(L), \text{ddf}, 0, \alpha\}]$$

**The two parts of the inequality identify two distributions for F**, each with its own denominator degrees of freedom (ddf).

**The left side of the inequality** defines a non-central F distribution (one whose non-centrality parameter, λ, is greater than zero) associated with a specific alternative hypothesis Ha.

**The right side of the inequality** invokes a central F distribution (whose non-centrality parameter equals 0), that is associated with the null hypothesis H0. It defines the "critical value" for the F distribution assumed under H0. At this critical value, the cumulative distribution function (CDF) of the central F-distribution, defined by numerator (ndf) and denominator degrees of freedom (ddf), is equal to α.

If the alternative hypothesis, embodied in the exemplary dataset, generates a non-central F statistic that is more extreme than the critical value for F defined under H0, we reject the null hypothesis.

## *The non-centrality parameter*

Stroup (1999) defines the non-centrality parameter,

$$\lambda = (L`\beta)` [L` (X`V^{-1}X)^{-1} L]^{-1} (L`\beta) \qquad \text{(Stroup, 1999)}$$

where L is the contrast matrix that defines the hypothesis, β is a vector of parameter estimates generated by the GLMM, X is the design matrix specified in the GLMM, and V is the matrix of expected or assumed variances and covariances.

The magnitude of λ depends on



the magnitude of the expected effect, which is specified in the quadratic form created by the matrix product L`β; this is analogous to a sum of squared differences among a linear combination of means.

the number and size of groups, specified in the design matrix X;

the pattern of independence versus correlation among observations, which are specified in the variance matrix V.

The variance matrix V can be simple, specifying a common variance $\sigma^2$ among independent observations, or it can describe complex patterns of variances and covariances. The V matrix can accommodate variance structures that assumed to be heterogenous or hierarchical (clustered or correlated).

## *GLMM estimates of power*

The tutorial illustrates, for each of seven examples, how to specify an alternative hypothesis in terms an exemplary dataset whose values for means reflect those expected under the alternative hypothesis. It shows how to specify random effects that reflect the expected values of variances and covariances.

In each example, the GLMM produces test statistics, which are processed in a SAS data step to calculate $\lambda$, the non-centrality parameter that defines the F distribution associated with the alternative hypothesis. The data step also estimates power. Each example employs the same data step:

```
data power;
  set fstats;
  alpha=0.05;
  lambda=numdf*fvalue;              /*non-centrality parameter*/
  fcrit=finv(1-alpha, numdf, dendf, 0); /*critical value for F distrib under H0*/
  power= 1- probf (fcrit, numdf, dendf, lambda);
run;
```

The approach estimates power by comparing features of two F distributions, one defined under the alternative hypothesis (the blue curve in Figure A1) and the other defined by the null hypothesis (the red curve).



The figure illustrates, for example 4 in the tutorial, the inequality:

$$\text{Power} = \Pr[F\{\text{rank}(L), \text{ddf}, \lambda\} > F\{\text{rank}(L), \text{ddf}, 0, \alpha\}]$$

**Figure A1. Power estimated for example 4, a two-arm "cross-sectional" CRT in which outcomes are measured twice in independent groups**

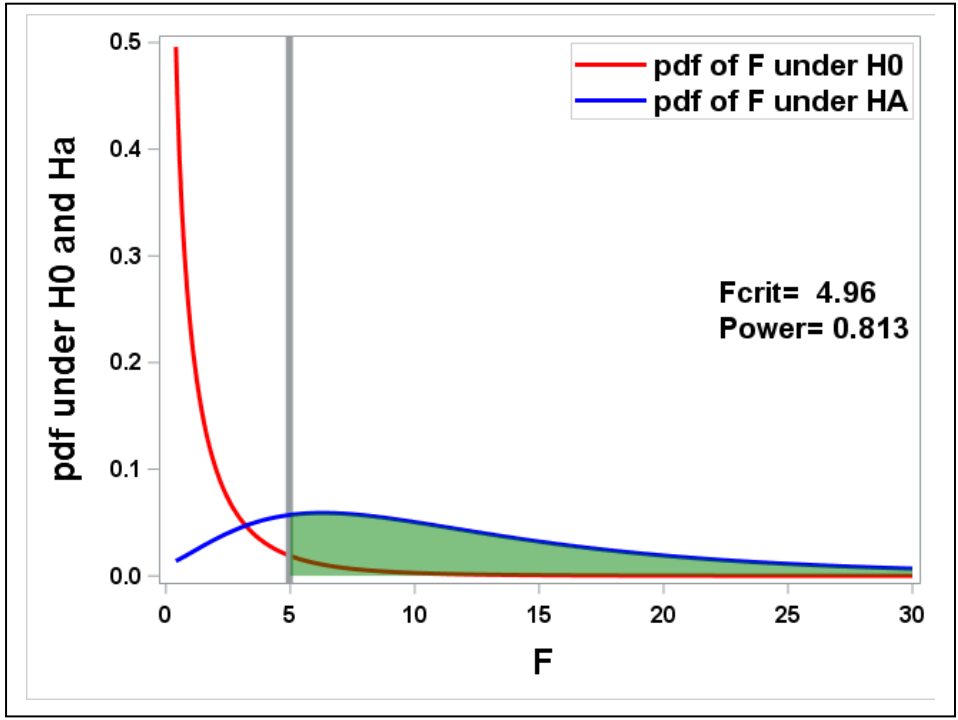

Power is the probability that, assuming the truth of a specific alternative hypothesis (a specific set of expected values for means and variances), one observes data that generate an F statistic equal to or more extreme than the critical value for F defined under the null hypothesis.